\documentclass[allclo]{FBSart}
\usepackage{amsfonts}
\usepackage{amssymb}
\usepackage[dvips]{graphicx}

\newcommand{\cd}{\makebox[0.08cm]{$\cdot$}}

\title{Critical stability of
three-body relativistic bound states with zero-range interaction}
\author{V.A. Karmanov$^{a,}$\thanks{\textit{E-mail address:}
karmanov@sci.lebedev.ru} and J.
Carbonell$^{b,}$\thanks{\textit{E-mail address:}
carbonel@lpsc.in2p3.fr}} \institute{$^a$Lebedev Physical
Institute, Leninsky Prospekt 53, 119991 Moscow, Russia\\ $^b$LPSC,
53 Avenue des Martyrs, 38026 Grenoble, France}

\runningauthor{V.A. Karmanov and J. Carbonell}
\runningtitle{Critical stability of three-body relativistic bound
states...}
\begin{document}

\maketitle
\begin{abstract}
For  zero-range interaction providing a given mass $M_{2}$ of the
two-body bound state, the mass $M_3$ of the relativistic
three-body bound state is calculated. We have found that the 
three-body system exists only when $M_2$ is greater than a critical 
value $M_{c}$ ($\approx 1.43\,m$ for bosons and $\approx 1.35\,m$ 
for fermions, $m$ is the constituent mass).  For $M_2=M_{c}$ the 
mass $M_3$ turns into zero and for $M_2<M_{c}$ there is no solution 
with real value of $M_3$.
\end{abstract}

%%%%%%%%%%%%%%%%%%%%%%%%%%%%%%%%%%%%%%%%%%%%%%%%%%%%
\section{Introduction}\label{intr}
Zero-range two-body interaction provides an important limiting
case which qualitatively reflects the characteristic properties of
nuclear and atomic few-body systems. In the nonrelativistic
three-body system it results in the Thomas collapse \cite{thomas}.
The latter means that the three-body binding energy tends to
$-\infty$, when the interaction radius tends to zero.

When the binding energy or the exchanged particle mass is not
negligible in comparison to the constituent masses, the
nonrelativistic treatment becomes invalid and must be replaced by
a relativistic one. Two-body calculations show that in the scalar
case, relativistic effects are repulsive (see e.g.
\cite{MC_PLB_00}). Relativistic three-body calculations with
zero-range interaction have been performed in a  minimal
relativistic model \cite{noyes} and in the framework of the
Light-Front Dynamics \cite{tobias}. It was concluded that, due to
relativistic repulsion, the three-body binding energy remains
finite and the Thomas collapse is consequently avoided. However,
in these works a cutoff was implicitly introduced. Because of
that, it was not clear, to what degree the finite binding energy
results from the relativistic repulsion, and to what degree --
from the cutoff. The latter can be imposed by many ways and it is
evident in advance that one can always find an enough strong
cutoff making the binding energy finite. Therefore we are
interested in a net effect of relativistic zero range interaction,
without any cutoff.

We present here our solution  \cite{ck03}  of the problem of three
equal mass ($m$) bosons interacting via zero-range forces.  In
addition, we consider also the three-fermion system. We  show that
the existence of the three-body system depends on the strength
two-body interaction. For strong enough interaction, instead of
the Thomas collapse its relativistic  counterpart takes  place.
Namely, when the two-body bound state mass $M_2$ decreases,  the
mass $M_3$ of the three-body system decreases as well and vanishes
at some critical value of $M_{2}=M_{c}$ ($\approx 1.43\,m$ for
three bosons and  $\approx 1.35\,m$ for three fermions). For
$M_{2}<M_{c}$ there are no solutions with real value of $M_3$,
what means -- from physical point of view -- that the three-body
system no longer exists.

%%%%%%%%%%%%%%%%%%%%%%%%%%%%%%%%%%%%%%%%%
\section{Three-boson system}\label{equat}
We use the explicitly covariant formulation of the Light-Front
Dynamics (see for a review \cite{cdkm}). The wave function is
defined on the light-front plane given by the equation $\omega\cd
x=0$, where $\omega$ is a four-vector with $\omega^2=0$,
determining the light-front orientation. In the particular case
$\omega=(1,0,0,-1)$ we recover the standard approach.

The three-body equation we consider is written for the vertex
function $\Gamma$, related to the wave function $\psi$:
\[\psi(1,2,3)=\frac{\Gamma(1,2,3)}
{M_0^2-M^2_3},\quad  M_0^2=(k_1+k_2+k_3)^2, \] where $M_3$ is the
three-body bound state mass.

The Faddeev amplitudes $\Gamma_{ij}$ are introduced in the
standard way:
$$\Gamma(1,2,3)=\Gamma_{12}(1,2,3)+\Gamma_{23}(1,2,3)+\Gamma_{31}(1,2,3)
$$ and one obtains a system of three coupled equations for them.
With the symmetry relations
$\Gamma_{23}(1,2,3)=\Gamma_{12}(2,3,1)$ and
$\Gamma_{31}(1,2,3)=\Gamma_{12}(3,1,2)$, the system is reduced to
a single equation for one of the amplitudes, say $\Gamma_{12}$.

For zero-range forces, the interaction kernel in momentum space is
replaced by a constant $\lambda$. This is precise meaning of the
relativistic zero-range interaction. For a given two-body bound
state mass $M_2$ the constant $\lambda$ is expressed through $M_2$
and disappears from the problem.

Equation for $\Gamma_{12}$ can be  rewritten in variables
$\vec{R}_{i\perp},x_i,$ ($i=1,2,3$), where $\vec{R}_{i\perp}$ is
the spatial component of the four-vector $R_i=k_i-x_ip$ orthogonal
to $\vec{\omega}$ and $x_i={\omega\cd k_i\over\omega\cd p}$
\cite{cdkm}. In general, $\Gamma_{12}$ depends on all  variables
($\vec{R}_{i\perp},x_i$), constrained by the relations
$\vec{R}_{1\perp}+\vec{R}_{2\perp}+\vec{R}_{3\perp}=0$,
$x_1+x_2+x_3=1$, but for  contact kernel it depends only on
$(\vec{R}_{3\perp},x_3)$ \cite{tobias}. The equation for the
Faddeev amplitude reads:
\begin{equation}\label{eq1}
\Gamma_{12}(R_{\perp},x)
=F(M_{12})\frac{\displaystyle{1}}{\displaystyle{(2\pi)^3}}
\displaystyle{\int_0^1} \displaystyle{dx'}
\displaystyle{\int_0^{\infty}}
\frac{\Gamma_{12}\left(R'_{\perp},x'(1-x)\right)\;d^2R'_{\perp}}
{\displaystyle{(\vec{R'}_{\perp}-x'\vec{R}_{\perp})^2+m^2-x'(1-x')M_{12}^2}}.
\end{equation}

The factor $F(M_{12})$ is the two-body off-shell scattering
amplitude. It corresponds to the fixed two-body bound state mass
$M_2$ and depends on the off-shell two-body effective mass
$M_{12}$. For $0\leq M_{12}^2<4m^2$ the calculation gives: $$
F(M_{12})=\frac{8\pi^2}{\frac{\displaystyle{\arctan y_{M_{12}}}}
{\displaystyle{y_{M_{12}}}} -\frac{\displaystyle{\arctan
y_{M_{2}}}}{\displaystyle{y_{M_{2}}}}}, $$ where $
y_{M_{12}}=\frac{M_{12}}{\sqrt{4m^2-M_{12}^2}}$ and similarly for
$y_{M_{2}}$. If $M_{12}^2<0$, the amplitude obtains the form: $$
F(M_{12})=\frac{8\pi^2}{\frac{\displaystyle{1}}
{\displaystyle{2y'_{M_{12}}}}\log
\frac{\displaystyle{1+y'_{M_{12}}}} {\displaystyle{1-y'_{M_{12}}}}
-\frac{\displaystyle{\arctan y_{M_{2}}}}
{\displaystyle{y_{M_{2}}}}}, $$ where
$
y'_{M_{12}}=\frac{\sqrt{-M_{12}^2}}{\sqrt{4m^2-M_{12}^2}}.
$
It has the pole at $M_{12}=M_2$.

The two-body mass squared $M_{12}^2$ is expressed through the
three-body variables as: $$
M^2_{12}=(1-x)M_3^2-\frac{R_{\perp}^2+(1-x)m^2}{x}. $$ The
three-body mass $M_3$ enters the equation (\ref{eq1}) through
$M_{12}^2$.

The arguments of $\Gamma_{12}$ run  the values $0\leq
R_{\perp}< \infty$, $0\leq x \leq 1$. The variable $M_{12}^2$
becomes negative at $x\to 0$ (the square is understood in the
sense of Minkowsky metric). By a replacement of variables the
equation (\ref{eq1}) can be transformed to the form of equation
(11) from \cite{tobias} except for the integration limits. In the
papers \cite{noyes,tobias}  the  variable $M_{12}^2$ was
constrained by positive values, that strongly restricts the domain
of variables $R_{\perp},x$ and, in this way, introduces a cutoff.

Being interested in studying the zero-range interaction, we do not
cut  the variation domain of variables $R_{\perp},x$. As we will
see, this point turns out to be crucial for the appearance of the
relativistic collapse.

%%%%%%%%%%%%%%%%%%%%%%%%%%%%%%%%%%%%%%%%%
\section{Three-fermion system}
The zero-range two-fermion kernel   can be constructed using many
different spin couplings. Our main interest is the influence of
the antisymmetrization of the wave function which should be taken
into account for any kernel. Therefore we solve the problem for a
simplified kernel:
\begin{equation}\label{eq4d}
{\cal K}^{\sigma'_1\sigma'_2}_{\sigma_1\sigma_2}(1,2;1',2')=
\lambda\; \bar{K}_{\sigma_1\sigma_2}(1,2)
K^{\sigma'_1\sigma'_2}(1',2'),
\end{equation}
where we denote:
%\begin{eqnarray}
\begin{equation}\label{eq4b1}
\bar{K}_{\sigma_1\sigma_2}(1,2)=\frac{m}{\omega\cd (k_1+k_2)}
\left[\bar{u}_{\sigma_1}(k_1)i\hat{\omega} \gamma_5U_c
\bar{u}_{\sigma_2}(k_2) \right] 
\end{equation}
%
%\label{eq4b1}\\
%K^{\sigma'_1\sigma'_2}(1',2')&=&\frac{m}{\omega\cd (k'_1+k'_2)}
%\left[u^{\sigma'_2}(k'_2)U_c i\hat{\omega}
%\gamma_5u^{\sigma'_1}(k'_1)\right]. \label{eq4c1}
%\end{eqnarray}
and $K^{\sigma'_1\sigma'_2}(1',2')=
\bar{K}^{\dagger\sigma'_1\sigma'_2}(1',2').$
The matrix $\hat{\omega}=\gamma^{\mu}\omega_{\mu}$ appears in the
contact interaction of fermions \cite{cdkm}. This kernel is
factorized relative to initial and final states. The divergence of
two-body scattering amplitude (at fixed $\lambda$) is logarithmic.
At fixed value of the two-body mass $M_2$ the amplitude becomes
finite, like in the boson case. In nonrelativistic limit, kernel
(\ref{eq4d}) corresponds to interaction in the $^1S_0$ state only.

The equation for the Faddeev component is generalized for the
three-fermion case by adding the spin indices. Its solution has
the form: $$ \Gamma^{\sigma}_{\sigma_1\sigma_2\sigma_3}(1,2,3) =
\bar{K}_{\sigma_1\sigma_2}(1,2) G^{\sigma}_{\sigma_3}(3), $$ where
$\bar{K}_{\sigma_1\sigma_2}(1,2)$ is defined in (\ref{eq4b1}). It
is antisymmetric relative to permutation $1\leftrightarrow 2$,
whereas the sum of three Faddeev components is antisymmetric
relative to permutation of any pair.

The $2\times 2$-matrix $G^{\sigma}_{\sigma_3}(3)$ can be
decomposed as: $$ G^{\sigma}_{\sigma_3}(3)=g_1
\bar{u}_{\sigma_3}(k_3)S_1u^{\sigma}(p) +g_2
\bar{u}_{\sigma_3}(k_3)S_2u^{\sigma}(p) $$ with the basis matrices
$$ S_1=\left[2x_3-(m+x_3 M_3)\frac{\hat{\omega}}{\omega\cd
p}\right], \quad S_2=m\frac{\hat{\omega}}{\omega\cd p}. $$ We get
system of two equations for the scalar functions $g_{1,2}$. One of
the equations contains only $g_1$ and namely it determines the
three-fermion bound state mass $M_3$.

%%%%%%%%%%%%%%%%%%%%%%%%%%%%%%%%%%%%%%%%%%
\section{Numerical results}\label{results}
\begin{figure}[h]
\begin{center}
\includegraphics[width=0.5\textwidth]{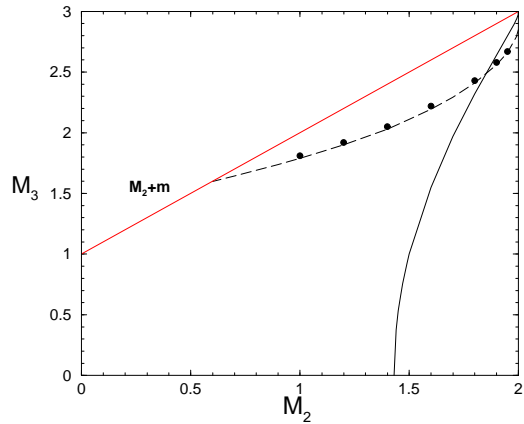}
\caption{Three-boson bound state mass $M_3$ versus the two-body
one $M_{2}$ (solid line). Results  obtained with integration
limits (\protect{\cite{tobias}}) are in dash line. Dots values are
taken from \cite{AMF_PRC_95}.}\label{Fig1}
\end{center}
\end{figure}

The results of solving equation (\ref{eq1}) for three bosons and
corresponding equation for three fermions are presented in what
follows. Calculations were carried out with constituent mass $m=1$
and correspond to the ground state. We represent in Fig.
\ref{Fig1} the three-body bound state mass $M_3$ as a function of
the two-body one $M_2$ (solid line) together with the dissociation
limit $M_3=M_2+m$.

Our results corresponding to integration limits \cite{tobias} are
included in Fig. \ref{Fig1} (dash line) for comparison. Values
from \cite{AMF_PRC_95} (corrected relative to \cite{tobias}) are
indicated by dots. In both cases the three-body binding energy is
finite and the Thomas collaps is absent, like it was already found
in \cite{noyes}. However, except for the zero binding limit, solid
and dash curves strongly differ from each other. In the two-body
zero  binding limit, the three-boson binding energy (solid line)
is $B_3\approx 0.012\,m$.

\begin{figure}[!ht]
\begin{center}
\includegraphics[width=0.5\textwidth]{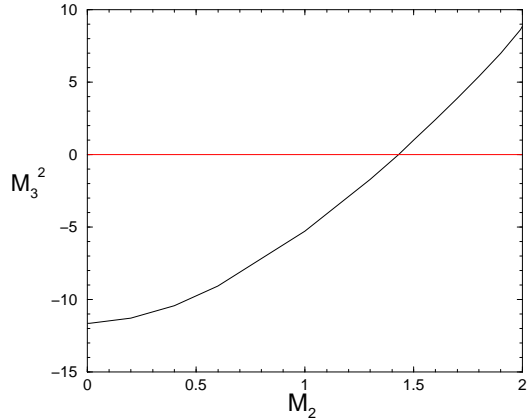}
\caption{Three-boson bound state mass squared $M_3^2$ versus
$M_{2}$.} \label{Fig2}
\end{center}
\end{figure}

When $M_2$ decreases, the three-body mass $M_3$  decreases very
quickly and vanishes at the two-body mass value
$M_{2}=M_{c}\approx 1.43$. This result was reproduced in Ref. 
\cite{bmfw}. Whereas the meaning of collapse as used
in the Thomas paper implies unbounded nonrelativistic binding
energies and cannot be used here, the zero bound state mass
$M_3=0$ constitutes its relativistic counterpart. Indeed, for
two-body masses below the critical value $M_c$, the three-body
system  no longer exists.

We would like to remark that for $M_2\leq M_c$, equation
(\ref{eq1}) posses square integrable solutions with negative
values of $M_3^2$. They have no physical meaning but $M_3^2$
remains finite in all the two-body mass range $M_2\in[0,2]$. The
results of $M_3^2$ are given in Fig. \ref{Fig2}. When $M_{2}\to
0$, $M_3^2$ tends to $\approx -11.6$.

\begin{figure}[!ht]
\begin{center}
\includegraphics[width=0.5\textwidth]{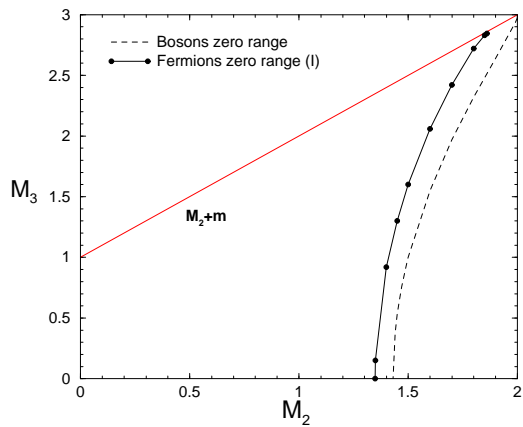}
\caption{Three-fermion bound state mass $M_3$ versus $M_{2}$
(solid line) in comparison to the three-boson bound state mass
(dash line).} \label{Fig3}
\end{center}
\end{figure}

The results for three-fermion system are shown in Fig. \ref{Fig3}.
Qualitatively they are similar to the three-boson case with the
curve shifted to  smaller $M_2$ values. As a consequence, the
critical value is $M_c\approx 1.35 $ instead of $1.43$ for bosons.
This value may however depend on the particular type of  spin
coupling used in the two-body kernel. Contrary to the boson case,
the three-fermion system is unbound in the two-body zero  binding
limit. The binding appears when the two-fermion system is already
bound by $B_2=0.1$, that is, for an interaction strong enough to
compensate the Pauli repulsion.

%%%%%%%%%%%%%%%%%%%%%%%%%%%%%%%%%
\section{Conclusion}
 In summary, we have considered the relativistic problem of
three equal-mass bosons and fermions, interacting via zero-range
forces constrained to provide finite two-body mass $M_{2}$. The
Light-Front Dynamics equations have been derived and solved
numerically.

We have found that the three-body bound state exists for two-body
mass values in the range $M_c\approx 1.43\,m \leq M_{2}\leq 2\,m$
for bosons and $M_c\approx 1.35\,m \leq M_{2}\leq 1.9\,m$ for
fermions. The Thomas collapse is avoided in the sense that
three-body mass $M_3$ is finite, in agreement with
\cite{noyes,tobias}. However, another kind of catastrophe happens.
Removing infinite binding energies, the relativistic dynamics
generates zero three-body mass $M_3$ at a critical value
$M_2=M_c$. For stronger interaction, i.e. when $0\leq M_{2}< M_c$,
there are no physical solutions with real value of $M_3$. In this
domain, $M_3^2$ becomes negative and the three-body system cannot
be described by zero range forces, as it happens in
nonrelativistic dynamics. This fact can be interpreted as a
relativistic collapse.

\begin{acknowledge}
This work is partially supported by the French-Russian PICS and
RFBR grants Nos. 1172 and 01-02-22002. Numerical calculations were
performed at CGCV (CEA Grenoble) and  IDRIS (CNRS).
\end{acknowledge}

%%%%%%%%%%%%%%%%%%%%%%%%%%%%%%%%%%%%%%%%%

\end{document}